\documentclass[prb,twocolumn,draft,amsmath,showpacs]{revtex4}
\usepackage{graphics}

\begin{document}

\bibliographystyle{prsty}
\input epsf

\title {Magneto-optical Kerr effect in $Eu_{1-x}Ca_{x}B_{6}$}

\author {G. Caimi, S. Broderick and H.R. Ott}
\affiliation{Laboratorium f\"ur Festk\"orperphysik, ETH Z\"urich,
CH-8093 Z\"urich, Switzerland}\

\author {L. Degiorgi}
\affiliation{Paul Scherrer Institute, CH-5232 Villigen and 
Laboratorium f\"ur Festk\"orperphysik, ETH Z\"urich,
CH-8093 Z\"urich, Switzerland}\

\author {A.D. Bianchi and Z. Fisk}
\affiliation{NHMFL-FSU, Tallahassee FL 32306, U.S.A.}

\date{\today}

\begin{abstract}
We have measured the magneto-optical Kerr rotation of ferromagnetic 
$Eu_{1-x}Ca_{x}B_{6}$ with x=0.2 and 0.4, as well as of $YbB_{6}$ 
serving as the non-magnetic reference material. As previously for 
$EuB_{6}$, we could identify a 
feature at 1 $eV$ in the Kerr response which is related with 
electronic 
transitions involving the localized 4f electron states. The absence 
of 
this feature in the data for $YbB_{6}$ confirms 
the relevance of the partially occupied 4f states in shaping the 
magneto-optical features of $Eu$-based hexaborides. Disorder by 
$Ca$-doping 
broadens the itinerant charge carrier contribution to the 
magneto-optical spectra. 
\end{abstract}
\maketitle

Recent investigations of the magneto-optical (MO) properties of 
$EuB_{6}$, a 
ferromagnet with a Curie temperature $T_{C}$ of 15.5 $K$, as a 
function of 
temperature and magnetic field, provided a variety of
results which, in particular, revealed an intimate link between the 
electronic 
properties and the bulk magnetization \cite{ref1,ref2,ref3}. Most 
informative in this respect were results from measurements of the 
polar Kerr 
rotation $\theta_{K}$, which is directly related to variations of the 
electronic excitation spectrum upon spontaneous and field induced 
magnetic 
order. Kerr rotation spectroscopy also yields information on 
electronic 
transitions involving both localized electronic orbits as well as 
itinerant charge carriers. The $\theta_{K}$ 
spectra of $EuB_{6}$ exhibit two particular features \cite{ref3}. The 
first, 
situated in the infrared spectral range at approximately 0.3 $eV$, is 
a resonance 
which increases in magnitude and shifts to higher energies with 
either decreasing temperature or increasing magnetic field. At all 
temperatures and fields this resonance coincides with the plasma edge 
feature of the 
optical reflectivity $R(\omega)$ (Refs. \onlinecite{ref1} and 
\onlinecite{ref2}) and is giant at low temperatures and high 
magnetic fields, i.e., of about 11 degree of rotation at 2 $K$ and 7 
$T$ (Ref. \onlinecite{ref3}). The second feature in 
$\theta_{K}(\omega)$ is a signal at 1 
$eV$, which also grows with decreasing temperature or increasing 
magnetic field, but does not shift in energy \cite{ref3}.

The resonance at 0.3 $eV$ is caused by the response of the itinerant 
charge 
carriers to magnetism, while the Kerr resonance at 1 $eV$ is 
associated with the 
MO response of the 4f-5d interband transitions 
\cite{ref3,ref4}. A phenomenological analysis, based on the classical 
Lorentz-Drude
dispersion theory adapted to the Kerr rotation phenomenon 
\cite{ref4}, 
revealed that the giant Kerr rotation pinned to the reflectivity 
plasma edge is indeed 
the consequence of an interplay between the response of the localized 
4f-states and the itinerant charge carriers, a suggestion  
previously put forward by Feil and Haas \cite{ref5,ref51}. The polar 
Kerr rotation 
at 1 $eV$ correlates directly with the bulk 
magnetization $M(H)$, confirming that the magnetic 
moments are essentially due to the localized f-electrons 
of the $Eu^{2+}$ ions \cite{ref3,ref4,ref6}. It is thus clear that 
Kerr 
spectroscopy allows for distinguishing between the roles of itinerant 
and localized 
electrons with respect to magnetic properties of a metal.

In order to confirm this scenario and its interpretation, we have 
repeated this type of experiments on material of the 
$Eu_{1-x}Ca_{x}B_{6}$ series and on $YbB_{6}$. It is well established 
that replacing $Eu$ in $EuB_{6}$ by isoelectronic $Ca$ changes both 
the 
magnetic and the transport properties of these hexaborides 
\cite{ref7,paschen}. Not surprisingly, the Curie temperature $T_{C}$ 
decreases significantly 
with increasing $Ca$ content. $YbB_{6}$ was chosen because it 
provides the possibility to probe the MO response of an 
essentially non magnetic hexaboride with divalent cations. In 
$YbB_{6}$, $Yb$ enters in its divalent configuration with all the 
14 f orbits occupied and thus without ionic magnetic moment 
\cite{ref8}.

\begin{figure}[t]
  \begin{center}
   \leavevmode
   \epsfxsize=.8\columnwidth \epsfbox {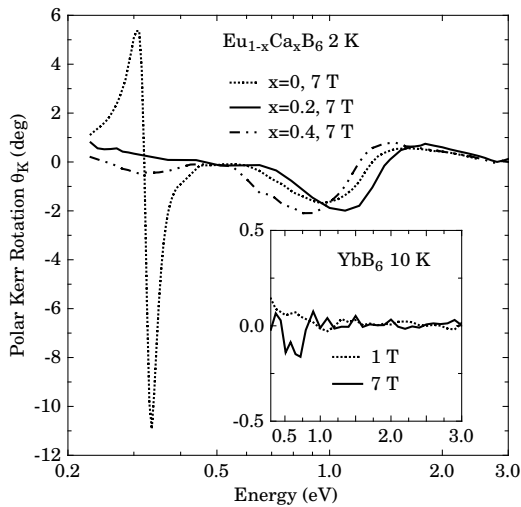}
    \caption{Polar Kerr rotation at 2 $K$ and 7 $T$ for 
    $Eu_{1-x}Ca_{x}B_{6}$. The inset displays the 1 and 7 $T$ Kerr 
spectra 
    of $YbB_{6}$ at 10 $K$. Note the much stretched scale for 
$\theta_{K}$ in the 
    inset.}
\label{Kerr}
\end{center}
\end{figure}

In our measurements of the Kerr rotation in the polar configuration, 
i.e., with the light beam oriented parallel to the magnetic field, 
linearly polarized light, 
a superposition of left (LCP) and right (RCP) circularly polarized 
components of equal magnitude, is reflected from the sample at a 
given temperature between 2 and 10 $K$ and in external magnetic 
fields from 0 to 10 $T$ (Refs. \onlinecite{ref3} and 
\onlinecite{refsam}). The polarization plane of the reflected 
light, with respect to that of the incident light, is rotated by a 
field 
and temperature dependent angle $\theta_{K}$. This is a consequence 
of 
the differing absorptions for the two circular polarizations. The 
Kerr 
angle is given by
\begin{equation}    
\theta_{K}=-\frac{1}{2}\Big(\Delta_{+}-\Delta_{-}\Big)=-Im\Big(\frac{\tilde{n}_{+}-\tilde{n}_{-}}{\tilde{n}_{+}\tilde{n}_{-}-1}\Big)
\end{equation}
where $\Delta_{\pm}$ are the phases of the complex reflectance 
$r_{\pm}(\omega)=\rho_{\pm}e^{i\Delta_{\pm}}$ and 
$\tilde{n}_{\pm}=n_{\pm}-ik_{\pm}$ are the 
complex refraction indices for LCP ($-$) and RCP (+) light, 
respectively 
\cite{ref3,ref4,refsam}. The relation between $\theta_{K}$ and the 
refraction indices in eq. (1) is strictly valid only for Kerr 
rotations not exceeding 15 degrees, which is the case here 
\cite{refsam}.

The single-crystalline samples of $Eu_{1-x}Ca_{x}B_{6}$, as well as 
$YbB_{6}$ were prepared by solution growth from $Al$ flux, using the 
necessary high-purity 
elements as starting materials. All samples were characterized by 
X-ray diffraction, dc-transport and thermodynamic 
experiments probing thermal properties \cite{ref7}. From 
magneto-transport data on $Ca$-doped 
$EuB_{6}$, the presence of $Al$ inclusions in 
the samples can be excluded \cite{ref7}. We have investigated two 
specimens with 
20$\%$ and 40$\%$ $Ca$-content and Curie temperatures $T_{C}$ of $5.3 
~K$ and 4.5 $K$, 
respectively \cite{ref7,paschen}.

\begin{figure}[t]
  \begin{center}
   \leavevmode
   \epsfxsize=.8\columnwidth \epsfbox {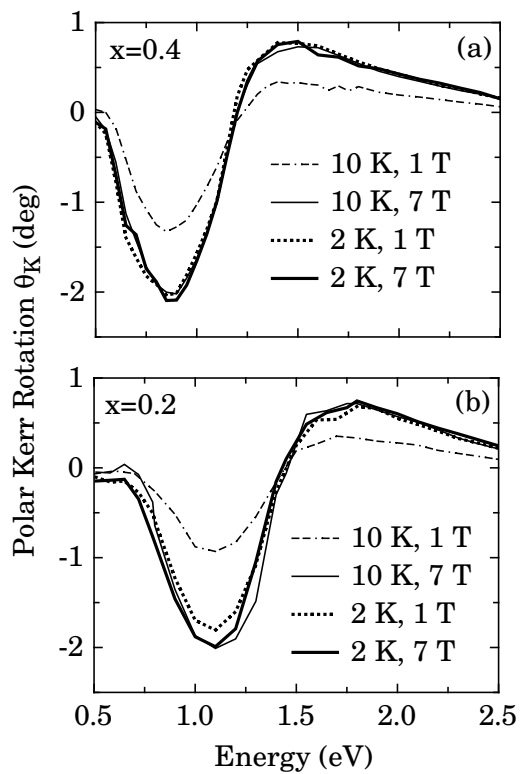}
    \caption{Temperature and magnetic field dependence of the polar 
Kerr 
    rotation $\theta_{K}(\omega)$ for $Eu_{1-x}Ca_{x}B_{6}$ in the 
spectral range around 1 
    $eV$.}
\label{Kerr1eV}
\end{center}
\end{figure}

Figure 1 summarizes $\theta_{K}(\omega)$ at 2 $K$ and 7 $T$ for 
20$\%$ and 40$\%$ 
$Ca$-doping. For the purpose of comparison we also display 
$\theta_{K}(\omega)$ for $EuB_{6}$ 
(Ref. \onlinecite{ref3}), measured under the same experimental 
conditions. 
The most obvious observation is that by $Ca$-doping the huge 
resonance at 
about 0.3 $eV$ is essentially wiped out, whereas the 
feature at about 1 $eV$ retains its shape and amplitude and shifts 
only 
slightly in energy. This feature in $\theta_{K}(\omega)$ is absent in 
the 
results \cite{comment} obtained for $YbB_{6}$, 
as may be seen in the inset of Fig. 1. Figure 2 emphasizes the field 
and temperature 
dependence of the 1 $eV$ resonance of the $Eu$-based compounds with 
x=0.2 and 0.4. The general trend is the 
same for both samples and bears similarities with the results 
\cite{ref3,ref4} for x=0. The 1 $eV$ feature increases in magnitude 
with 
decreasing temperature and increasing field, but does not exhibit a 
shift of the resonance frequency. The increase in magnitude tends to 
saturate at low temperatures and high magnetic fields.

Magneto-optical reflectivity data, to be presented and discussed 
elsewhere \cite{ref9}, indicate that $Ca$-doping induces an 
enhancement of the 
scattering rate for itinerant charge carriers, in accordance with dc 
transport results \cite{ref7}. This in turn leads to a so-called 
overdamped behaviour of 
the reflectivity, particularly at high temperatures and low fields. 
The sluggish onset of the $R(\omega)$ plasma edge behaviour in 
$Ca$-doped 
$EuB_{6}$ is apparent in Fig. 3b, where $R(\omega)$ at 10 $K$ 
and 7 $T$ for the binary compound \cite{ref3} is also shown for 
comparison.   
The $R(\omega)$ plasma edge in $Eu_{1-x}Ca_{x}B_{6}$ for $x\ne 0$ 
extends over some 
range in $\omega$ around 0.3 $eV$ and 
the considerable blue-shift of the edge, as observed for $EuB_{6}$ 
(Refs. \onlinecite{ref1} and \onlinecite{ref2}) 
with decreasing temperature or increasing field, is absent 
\cite{ref9}.

\begin{figure}[t]
  \begin{center}
   \leavevmode
   \epsfxsize=.8\columnwidth \epsfbox {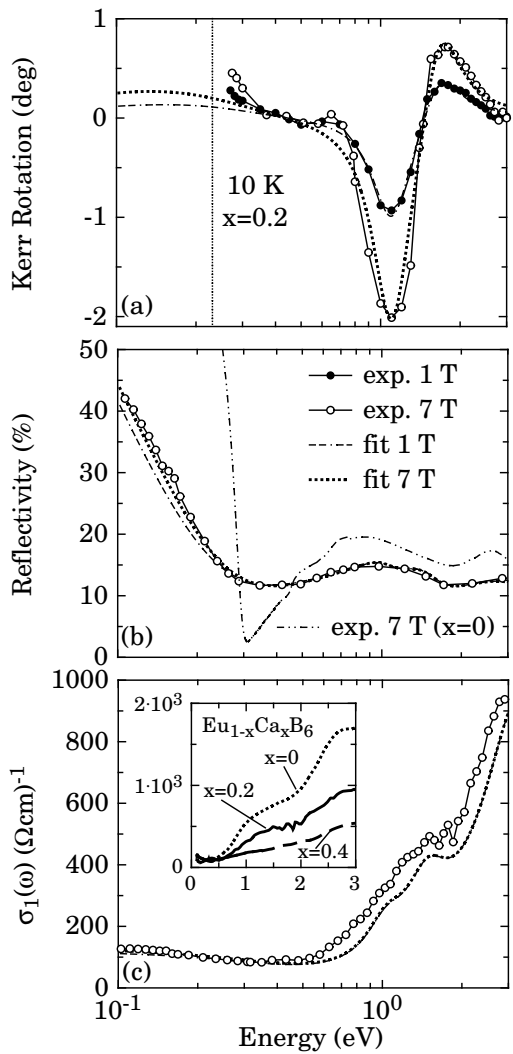}
    \caption{Comparison of the measured and calculated (Lorentz-Drude 
model) 
    Kerr rotation, reflectivity and real part $\sigma_{1}(\omega)$ of 
the optical conductivity 
    at 10 $K$ and selected fields for $Eu_{0.8}Ca_{0.2}B_{6}$. The 
low energy limit of the Kerr spectrometer (i.e., 0.23 $eV$) 
is marked with a vertical thin dotted line in panel (a). For the
    purpose of clarity, the panels (b) and (c) show only the 
    7 $T$ data, since the field dependence is negligible in this 
    spectral range, as is confirmed by the corresponding calculated 
    curves. For comparison, the sharp onset of the 
    $R(\omega)$ plasma 
    edge of $EuB_{6}$ at 10 $K$ and 7 $T$ is also shown in panel (b). 
The inset in panel (c) shows $\sigma_{1}(\omega)$ 
    for $Eu_{1-x}Ca_{x}B_{6}$ in the spectral range around 1 $eV$. 
    For all panels the same symbols for x=0.2 as those shown in panel 
(b) apply.}
\label{fit}
\end{center}
\end{figure}

As we have recently demonstrated \cite{ref4,refsam}, the extended 
Lorentz-Drude model 
\cite{wooten} is 
quite successful, in reproducing the MO features of $EuB_{6}$. 
Therefore, for the analysis of the data presented here we have 
adopted the same 
approach as that employed in Ref. \onlinecite{ref4}. It turns out 
that for the samples of the $Eu_{1-x}Ca_{x}B_{6}$ series with $x\ne 
0$, two Drude components are 
necessary to account for the low frequency electrodynamic 
response \cite{ref9}. Only one of 
the Drude terms is really relevant for determining the dc 
($\omega\to 0$) transport properties and it also dominates the 
temperature and magnetic field dependences of 
$\sigma_{1}(\omega)$ at low frequencies. The second 
Drude-type contribution accounts for the spectral weight background 
in $\sigma_{1}(\omega)$. The scattering rate of the first Drude term is
rather small but temperature and field dependent, 
while the one of the second Drude term is large, and temperature 
and field independent \cite{ref9}. The essence of the present discussion 
is not really affected by the way by which the metallic component in the optical 
properties is fitted. For the 
reproduction of the optical data in the spectral range above 1 $eV$, which are dominated by 
interband transitions, four Lorentz harmonic oscillators (h.o.) had to be 
considered. The four h.o.'s are at 1 and 1.5 $eV$ for 20$\%$ 
$Ca$-doping, at 0.8 and 1.3 $eV$ for 40$\%$ $Ca$-doping, and at 3.4 
and $\sim 11 ~eV$ for both $Ca$-contents \cite{ref11}. The energy 
splitting 
of the h.o.'s around 1 $eV$ is consistent with the experimentally 
verified splitting of the f-states \cite{lang}.

Since above 2.5 $eV$ no temperature and 
magnetic field dependence of the optical properties were detected, 
the parameters of the Lorentz 
h.o.'s at 3.4 and 11 $eV$ were 
established at 10 $K$, in order to fit the measured reflectivity 
$R(\omega)$ as well as all the optical functions up 
to 12 $eV$. For the fits to the Kerr rotation data in the relevant 
energy 
interval (0.23-4 $eV$), these latter parameters were then fixed for 
all 
temperatures and magnetic fields. The parameters of the Drude 
components and 
of the two h.o.'s necessary for describing the 1 $eV$ feature in 
$\theta_{K}$ were, however, allowed to 
vary as a function of temperature and field \cite{ref11}. The overall 
features of the MO 
properties and particularly their temperature and magnetic field 
dependences can be reproduced 
quite accurately. Figure 3 exemplifies the fit quality for the sample 
with 20$\%$ $Ca$-content
at 10 $K$ and two selected fields. Similarly good fits (not shown 
here) were obtained \cite{refsam}
for $Eu_{1-x}Ca_{x}B_{6}$ with x=0.4. 
Figure 3 also demonstrates that the same set of parameters 
\cite{ref11} can equally well 
account for the fits of $\theta_{K}(\omega)$, the reflectivity 
$R(\omega)$ and 
the real part $\sigma_{1}(\omega)$ of the optical conductivity. 
Another information that may be obtained from Fig. 3 is the 
correspondence between 
the 1 $eV$ signal in $\theta_{K}(\omega)$ and the related interband 
absorption, manifested by the broad feature in $R(\omega)$ and a 
shoulder in $\sigma_{1}(\omega)$ at the same energy.

The reduced number of 4f states upon $Ca$-doping is manifest in the 
zero-field 
$\sigma_{1}(\omega)$ curves of $Eu_{1-x}Ca_{x}B_{6}$ in the spectral 
range 
around 1 $eV$. The shoulder in 
$\sigma_{1}(\omega)$ at 1 $eV$ (inset of Fig. 3c), associated with 
the 4f-5d interband transitions 
\cite{ref3,ref4,kimura}, decreases with $Ca$-doping. The total 
absence of the 1 $eV$ 
feature in $\theta_{K}(\omega)$ for the non-magnetic 
reference material $YbB_{6}$ (inset of Fig. 1) confirms the validity 
of the interpretation of this feature. A 
Kerr rotation is only encountered, if the absorption 
coefficients for LCP and RCP light are unequal. This
situation is not expected for
$YbB_{6}$ with fully occupied 4f electron shells of the $Yb^{2+}$ 
ions.  

The calculated $\theta_{K}(\omega)$ (Fig. 3a) exhibits no sharp 
resonance in the spectral range of the 
reflectivity plasma edge (Fig. 3b) and no blue-shift of the same, 
neither within nor below the experimentally accessible 
range for the measurements of $\theta_{K}(\omega)$ (thin dotted line 
in 
Fig. 3a). Thus it agrees with the experimentally observed 
$\theta_{K}(\omega)$.
In $EuB_{6}$ the sharp onset of the $R(\omega)$ plasma edge and its 
blue shift \cite{ref2} were unambiguously correlated with the 
$\theta_{K}$ resonance at about 0.3 $eV$ (Ref. \onlinecite{ref3}). 
This interpretation was supported by the Lorentz-Drude fit even for 
energies extending 
to below the experimental limit of the Kerr spectrometer 
\cite{ref3,ref4}. The overdamped behaviour of the $R(\omega)$ plasma 
edge of $Ca$-doped $EuB_{6}$ (Fig. 3b) 
suppresses the necessary resonance conditions, as postulated by Feil 
and 
Haas \cite{ref5}, for a giant response in $\theta_{K}(\omega)$. Based 
on 
magnetotransport data it was suggested that in $Ca$-doped $EuB_{6}$ 
the 
formation of magnetic domain walls is favored by the disorder on the 
cation sublattice \cite{ref7}. This is 
probably reflected in the large damping (scattering rate) of the 
itinerant charge 
carriers. Upon $Ca$-doping the sharp onset of the $R(\omega)$ 
plasma edge and the related Kerr resonance, as 
observed for $EuB_{6}$ (Refs. \onlinecite{ref2,ref3,ref4}), are 
broadened and wiped out, respectively. 

\begin{figure}[t]
  \begin{center}
   \leavevmode
   \epsfxsize=.8\columnwidth \epsfbox {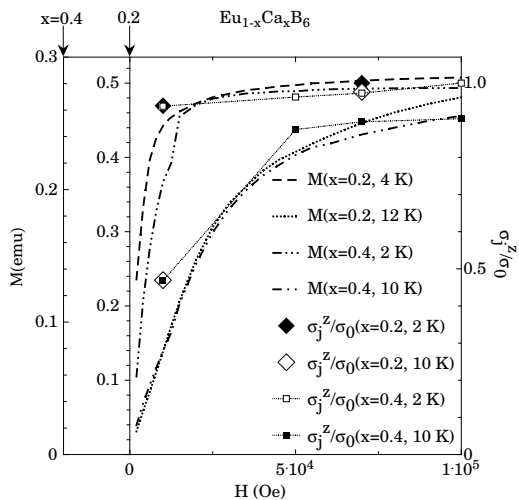}
    \caption{Comparison between the measured magnetizations M (Ref. 
    \onlinecite 
    {wigger}) and the quantity $\sigma_{j}^{z}/\sigma_{0}$ (eq. (2), 
with 
    $\sigma_{0}$ representing the corresponding $\sigma_{j}^{z}$ 
    value at the lowest temperature and highest 
    field) calculated for the h.o. around 1 $eV$ and appropriately 
scaled to $M(H)$ at selected 
    temperatures.}
\label{spin-pol}
\end{center}
\end{figure}

The relevant quantity of the Kerr spectroscopy, which is selectively 
sensitive to spin-polarized states, is the relative distribution 
of spectral weight, i.e., the unequal absorption coefficients, 
involved in the optical transitions for RCP and 
LCP light. It is parameterized by the so-called weight factors 
$f^{\pm}$ 
(Refs. ~\onlinecite{ref4} and \onlinecite{refsam}) which shape the 
dispersion-like 
curve of $\theta_{K}(\omega)$ at 1 $eV$. We have introduced the 
quantity
\begin{equation}
    \sigma_{j}^{z}=\vert (f^{+}-f^{-})/(f^{+}+f^{-})\vert=\vert 
f^{+}-1\vert, 
\end{equation}
with the condition that $f^{+}+f^{-}=2$, because of the sum rule 
constraints. For $EuB_{6}$ we have demonstrated \cite{ref4,refsam}
that $\sigma_{j}^{z}$ is phenomenologically related to the 
magnetization 
$M(H)$, thus acting as the MO 
counterpart of the moment polarization. Inspection of the $f^{\pm}$ 
factors \cite{ref11} for the h.o.'s describing the 1 $eV$ feature and 
a similar analysis in terms of 
$\sigma_{j}^{z}$ for the $Ca$-doped $EuB_{6}$ (Fig. 4) confirm that 
the 
1 $eV$ feature in $\theta_{K}(\omega)$ is again associated with the 
MO response 
of the 4f electron states which are also responsible for the 
magnetization 
of the system \cite{ref6}. In fact, 
$\sigma_{j}^{z}(H)$ mimics the trend of the magnetization (Fig. 4), 
just as for $EuB_{6}$ (Ref. \onlinecite{ref4}). As for $EuB_{6}$, the 
strongly scattered 
itinerant charge carriers in the $Ca$-doped compounds are hardly 
polarized (i.e. $f^{\pm}\simeq 1$), indicating once more 
their minor role in shaping the magnetization. The flat $\theta_{K}$ 
spectrum in $YbB_{6}$ at 1 $eV$ implies $f^{\pm}=1$ and 
consequently $\sigma_{j}^{z}=0$, known as a so-called diamagnetic 
situation 
with respect to the Kerr rotation \cite{refsam}.

Taken together, our MO data on $Eu_{1-x}Ca_{x}B_{6}$ and $YbB_{6}$ 
indicate one successful way for achieving large Kerr rotations. 
Essential are electronic excitations out of incompletely filled 4f 
electron orbitals. Their effect may significantly be enhanced by 
excitations of itinerant electrons but it is required that the 
corresponding scattering relaxation rate is low enough to ascertain a 
steep slope at the onset of the Drude plasma edge in the 
reflectivity, thus confirming an earlier suggestion of Feil and Haas 
\cite{ref5}.

\acknowledgments
The authors wish to thank A. Perucchi, G.A. Wigger and R. Monnier for 
fruitful discussions. This 
work has been supported by the Swiss National Foundation for the 
Scientific Research.

\newpage


\begin{thebibliography}{10}
\bibitem{comment}
At 7 T we observe a weak feature in $\theta_{K}(\omega)$ of $YbB_{6}$ 
at about 0.5 $eV$, which does not 
follow a consistent behaviour either in temperature or in field. We 
believe that it 
is due to scattering of the data.
\bibitem{wigger}
The magnetization has been measured up to 5.5 $T$ (Ref. 
\onlinecite{ref7}). By assuming a Brillouin function for the overall 
behaviour of 
$M(H,T)$, a simple fit of the measured magnetization data leads to 
the 
values of $M$ up to 10 $T$ at different temperatures.
\end{thebibliography}
\end{document}